\documentclass[final]{aipproc}

\layoutstyle{6x9}

\begin{document}

\title{Precision Measurements of the Proton Elastic Form Factor Ratio}
\author{D. W. Higinbotham}{
  address={Jefferson Lab, Newport News, VA 23606, USA}
}

\begin{abstract}
New high precision polarization measurements of the proton elastic form factor ratio in the  
Q$^2$ range from 0.3 to 0.7~[GeV/c]$^2$ have been made. 
These elastic H(e,e'p) measurements were done 
in Jefferson Lab's Hall A using 80$\%$ longitudinally polarized electrons 
and recoil polarimetry.  For Q$^2$ greater than 1~[GeV/c]$^2$, previous polarization 
data indicated a strong deviation of the form factor ratio from unity which sparked 
renewed theoretical and experimental interest in how two-photon diagrams have been 
taken into account.  The new high precision data indicate that the deviation 
from unity, while small, persists even at Q$^2$ less than 1~[GeV/c]$^2$. 
\end{abstract}

\maketitle

\section{Introduction}

Understanding the four-momentum transfer squared, Q$^2$, dependence 
of the proton's electro-magnetic form factors is fundamental to 
understanding the proton and as well
as a necessary input parameter in many calculations.
Cross section measurements generally show that the ratio
of the proton's electric to magnetic form factor is 
basically unity~\cite{Qattan:2004ht},
while at Q$^2 > 1$[GeV/c]$^2$ recoil polarization 
measurements~\cite{Jones:1999rz,Gayou:2001qd,Gayou:2001qt,Punjabi:2005wq} 
as well a beam-target polarization measurement~\cite{Jones:2006kf} have observed 
a deviation from unity.  At this time, the most likely cause 
for the difference between the cross section results and the polarization
results is the two-photon part of the radiative corrections~\cite{Guichon:2003qm}.

\section{Radiative Corrections}

\begin{figure}[t]
  \label{higinbotham:mo-tsai}
  \includegraphics[width=\textwidth]{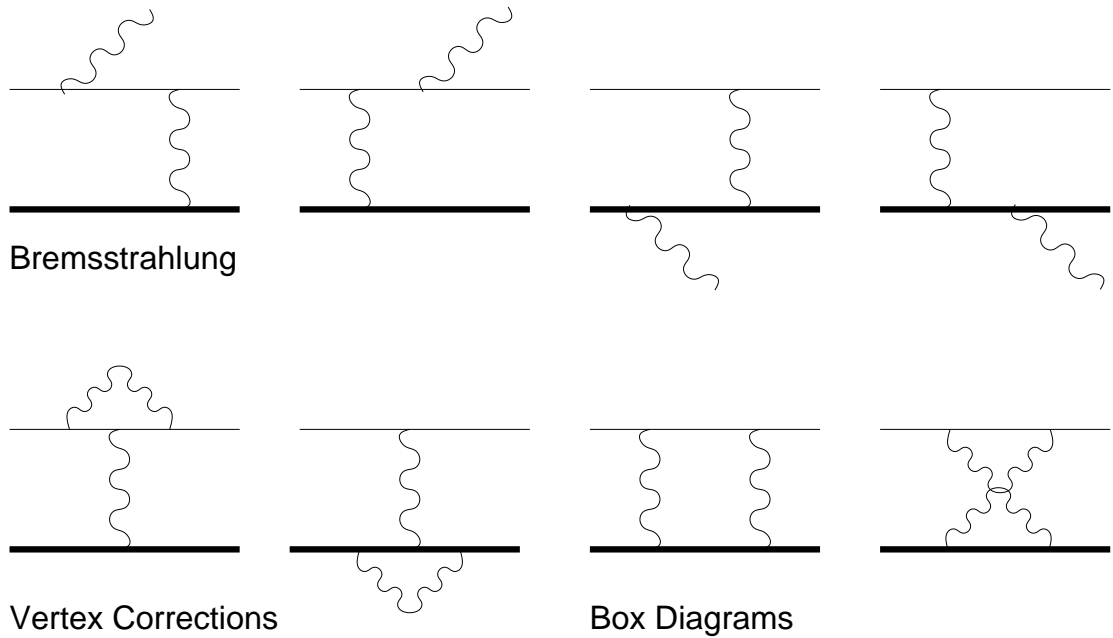}
  \caption{Shown are the standard Feynman diagrams included in the calculation
of radiative corrections in a lepton-hadron scattering experiment.  The thin
line represents the lepton, the thick line the hadron and the wavy line(s) the
virtual photon(s).
 }
\end{figure}

It is important to understand that two-photon diagrams have been included
in the standard cross section radiative corrections such as
Mo and Tsai~\cite{Mo:1968cg}, but they have been included with varying
degrees of approximation.   For example, the Feynman diagrams for 
Mo and Tsai's approach are shown in Fig.~\ref{higinbotham:mo-tsai} where
the proton's structure is neglected and the two-photons were 
only allowed to have either all the four-momentum transfer or none. 

\begin{figure}[t]
\label{higinbotham-box}
\centering
\includegraphics[width=0.8\textwidth]{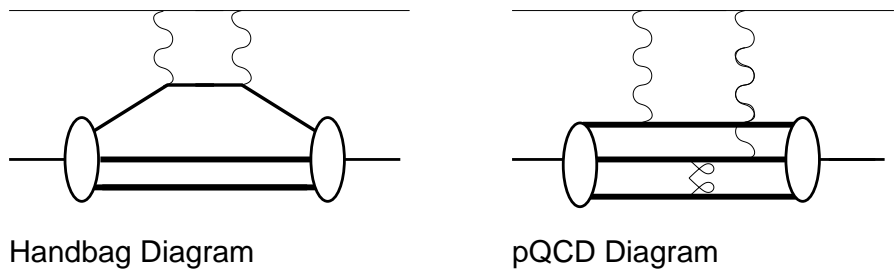}
\caption{Shown are the handbag and pQCD diagrams for handling the two-photon corrections in 
a more detailed picture.  In the handbag diagram the virtual photons couple to a single
parton, while in the pQCD diagram, the virtual photons couple to different partons with a gluon 
coupling the last two partons together.}
\end{figure}

Recent calculations of radiative corrections not only integrate over all possible
photon energies and the proton's structure, but even allow the proton within the box-diagrams
to be off-shell as shown in Fig.~\ref{higinbotham-box}.  These effects make the
calculations particularly challenging as it is the proton's structure
that one is trying to 
determine~\cite{Afanasev:2005mp,Blunden:2005ew,Borisyuk:2008db,Kivel:2009eg}
In fact, the unexpected discrepancy
in the cross section and asymmetry measurements should, in the long run,
dramatically improve our understanding of not only radiative corrections, but
also the proton's structure.

\section{Low Q$^2$ Form Factor Measurements}

The observed discrepancy between cross section and asymmetry measurements as well as the calculation
of the two-photon corrections has focused on Q$^2 > 1$~[GeV/c]$^2$.  Previous low Q$^2$ polarization 
measurements~\cite{Crawford:2006rz,Ron:2007vr} perhaps saw hints of an effect for Q$^2 < 1$~[GeV/c]$^2$, but
certainly nothing definitive.  
Experiment E08-007 at 
Jefferson Lab~\cite{E08-007} made a high precision survey of the form factor ratio in the Q$^2$ range from
0.3 to 0.7 [GeV/c]$^2$ using the recoil polarization technique.  The experiment was  done in 
Hall~A~\cite{Alcorn:2004sb} with a High Resolution Spectrometer (HRS) detecting the
elastically scattered proton and the BigBite magnet~\cite{deLange:1998qq,deLange:1998au} along with a lead glass calorimeter used
for tagging the elastic electrons.  In addition, a 6~cm long liquid hydrogen target was used  
along with 80$\%$ longitudinally polarized electrons
from the Continuous Electron Beam Accelerator Facility (CEBAF)~\cite{Leemann:2001}.

As knowledge of the spin precession within the high resolution spectrometer is critical for the extraction of 
the form factors, three different HRS momentum settings were used for each of the kinematics shown in 
Table~\ref{higinbotham-kinematics}.  
This was possible since the nominal momentum bite of the 
HRS is $\pm$ 4.5\%, so by choosing momenta of $0\%$ and $\pm2\%$ around the nominal elastic kinematic settings, we 
changed the spin precession while still staying well within the nominal acceptance of the device.  Each of these
settings was also measured to 1-2\% statistics so that systematic effects could be studied and not confused with statical
fluctuations.
It is worth noting that spin precession of spin-1/2 particles though a dipole
magnetic field is extremely well understood; it is possible to calculate the thousands of degrees of spin precession 
the polarized electrons at CEBAF undergo as they travel five times around the accelerator~\cite{Higinbotham:2009ze}. 
In order to also take into account the effects
of the super-conducting quadrupoles in the HRS, COSY, a spin transport modelling program, is used.  
For events with rays that pass near the central ray of the spectrometer, 
the COSY model and the simple dipole model agree.  Without using COSY, as one goes away from the central ray in angle, a
strong slope can be seen in the form factor ratio.  This slope goes away once the COSY spin matrix is applied.

The preliminary results of this experiment indicate that the deviation of the ratio of the form factor smoothly continues
into the low Q$^2$ and that there is no sharp transition to unity around Q$^2$ equal to one.  It does appear that a rapid
change in the ratio must occur for the Q$^2$ less then 0.3~[GeV/c]$^2$ either with a rise to zero such as in relativistic
pQCD models~\cite{demelo} or with a rise above one near Q$^2$ of 0.1~[GeV/c]$^2$ 
and then a return to unity at a Q$^2$ of zero.
 
\begin{table}[h]
\begin{tabular}{ccccc}
\hline
\tablehead{1}{c}{b}{$Q^2$ $[$GeV/c$]^2$} &
\tablehead{1}{c}{b}{$\theta_{p}$ $[$deg$]$} &
\tablehead{1}{c}{b}{$P_{p}$ $[$GeV/c$]$} &
\tablehead{1}{c}{b}{$\theta_{e}$ $[$deg$]$} &
\tablehead{1}{c}{b}{$P_{e}$ $[$GeV/c$]$} \\
\hline
0.30 & 60.0 & 0.565 & 30.0 & 1.03 \\
0.35 & 57.5 & 0.616 & 30.0 & 1.01 \\
0.41 & 55.0 & 0.668 & 35.0 & 0.978 \\
0.45 & 53.0 & 0.710 & 35.0 & 0.954 \\
0.50 & 51.0 & 0.752 & 40.0 & 0.928 \\
0.55 & 49.0 & 0.794 & 40.0 & 0.901 \\
0.60 & 47.0 & 0.836 & 45.0 & 0.874 \\
0.70 & 43.5 & 0.913 & 50.0 & 0.823 \\ \hline
\end{tabular}
\label{higinbotham-kinematics}
\caption{Shown are the central angles and momentum for the high resolution spectrometer used for 
proton detection and the kinematics for the BigBite spectrometer used for electron detection.
For each $Q^2$, three different spectrometer momentum settings were used;
all within the nominal momentum acceptance of the spectrometer.}
\end{table}

\section{Nucleons in Deuterium}

The simplest system in which to study a bound nucleon is the deuteron.  Polarized beam and vector polarized target experiments
at NIKHEF showed that at low Q$^2$ the D(e,e'p)n reaction asymmetry was the same as the H(e,e'p) elastic asymmetry and only
at missing momentum greater than approximately $100$~[MeV/c] was an appreciable deviation observed~\cite{Passchier:2001uc}.  
The MIT-Bates recoil polarization experiment also didn't see any large difference between hydrogen elastic and low missing
momentum deuteron quasi-elastic scattering~\cite{PhysRevLett.80.452,PhysRevLett.82.2221}.
This seems to contradict the recoil polarization results of B.~Hu {\it et al.}~\cite{Hu:2006fy} where a deviation was reported.  
Since the data of B.~Hu was taken with the same equipment as the new high precision hydrogen data, it is straightforward
to include the new results and look at the ratios again as shown in Table~\ref{higinbotham-hu}.  
By doing this, the $\chi^2/n$ of the hypothesis of a ratio of one goes 
from 1.7 to 0.7.  Thus, within experimental uncertainties, the low missing momentum data are consistent 
with no effect.  It is now clear that what drove the original
deviation was a high free proton form factor ratio.  The current result again validates
the common technique of extracting neutron properties from reactions on
neutrons in a deuteron target, with small theoretical corrections. 

\begin{table}[ht]
\begin{tabular}{cccccc}
\hline 
\tablehead{1}{c}{b}{$Q^2$\\$[$GeV/c$]^2$} &
\tablehead{1}{c}{b}{Deuteron\\ $\mu G_E/G_M$} &
\tablehead{1}{c}{b}{Proton \\  $\mu G_E/G_M$} &
\tablehead{1}{c}{b}{Old Ratios\\$P_D/P_{free}$} &
\tablehead{1}{c}{b}{New Proton\\$\mu G_E/G_M$} &
\tablehead{1}{c}{b}{New Ratio\\$P_D/P_{free}$} \\ \hline
0.43	& 0.92 $\pm$ 0.03  & 0.99 $\pm$ 0.03 & 0.92 $\pm$ 0.04 & 0.93 $\pm$ 0.01 & 0.99 $\pm$ 0.03 \\
1.00 & 0.88 $\pm$ 0.01  & 0.88 $\pm$ 0.02 & 1.00 $\pm$ 0.03 & & \\
1.61 & 0.93 $\pm$ 0.04  & 0.87 $\pm$ 0.04 & 1.08 $\pm$ 0.07 & & \\ 
\hline
\end{tabular}
\label{higinbotham-hu}
\caption{Shown is the results from B. Hu~\cite{Hu:2006fy} along with the new proton form factor data from E08-007.
By including the new data, the $\chi^2/n$ of a flat line fit of the ratio goes from 1.7 to 0.7; thus indicating
the low $P_m$ (e,e'p) reaction from deuterium is consistent with the free proton case.  Do note that for the
high $Q^2$ point of 1.61, there are large systematics errors in determining the individual ratios that cancel in the ratio.} 
\end{table}

\section{Future Experiments}

It would be very interesting to push high precision proton form factor measurements to
lower Q$^2$, but the technique of recoil polarization requires 
protons with reasonable momenta, so it is not feasible to go to lower Q$^2$ with the
setup described herein.
With an electron-proton collider, it would be possible to get to lower
Q$^2$ due to the boost of the proton's momentum, or one can do the measurement with
a polarized beam and a fixed target where only the scattered electron needs to be detected.
Such an experiment is planned for
2012 at Jefferson Lab as the second half of experiment E08-007~\cite{E08-007} and will cover 
the Q$^2$ range from 0.015 to 0.4~[GeV/c]$^2$. 

If two-photon diagrams are truly the solution to the discrepancy between the cross section and the asymmetry
experiments, this should be clearly seen in the upcoming high precision electron-proton and 
positron-proton cross section ratio experiments
such as the Jefferson Lab's Hall~B experiment~\cite{E04-116} presented herein by B.~Raue (FIU),
the VEPP-III experiment~\cite{Arrington:2004hk} and the Olympus experiment at DESY~\cite{Kohl:2009zz}.


\begin{theacknowledgments}
I thank Carl Carlson (W\&M) for teaching me about the history of two-photon corrections
and Ph.D. student Xiaohui Zhan (MIT) for her outstanding work analyzing the new low Q$^2$ data.
This work was supported by the U.S.\ Department of Energy,
and Jefferson Science Associates which operates
the Thomas Jefferson National Accelerator Facility under DOE
contract DE-AC05-06OR23177.

\end{theacknowledgments}

\bibliographystyle{aipproc}   

\bibliography{Hadron2009-GEP}

\end{document}